\documentclass[apj]{emulateapj}

\usepackage{natbib}
\slugcomment{Mawet et al. 2014, accepted to ApJ}

\shorttitle{Contrast limitations set by small sample statistics}
\shortauthors{Mawet et al.}

\begin{document}


\title{Fundamental limitations of high contrast imaging\\
set by small sample statistics}


\author{D. Mawet, J. Milli, Z. Wahhaj}
\affil{European Southern Observatory, Alonso de Cord\'ova 3107, Vitacura, Santiago, Chile}


\author{D. Pelat}
\affil{LUTH, Observatoire de Paris, 5 pl. J. Janssen, F-92195 Meudon, France}

\author{O. Absil, C. Delacroix}
\affil{D\'epartement d'Astrophysique, G\'eophysique et Oc\'eanographie, Universit\'e de Li\`ege, 17 All\'ee du Six Ao\^ut, B-4000 Li\`ege, Belgium}


\author{A. Boccaletti}
\affil{LESIA, Observatoire de Paris, 5 pl. J. Janssen, F-92195 Meudon, France}


\author{M. Kasper}
\affil{European Southern Observatory Headquarters, Karl-Schwarzschild-Str. 2, 85748 Garching bei M\"{u}nchen, Germany}

\author{M. Kenworthy}
\affil{Leiden Observatory, Sterrewacht Leiden, P.O. Box 9513, Niels Bohrweg 2, 2300-RA Leiden,
The Netherlands}

\author{C. Marois}
\affil{NRC, Herzberg Institute of Astrophysics, Victoria, BC V9E 2E7, Canada}

\author{B. Mennesson}
\affil{Jet Propulsion Laboratory, California Institute of Technology, 4800 Oak Grove Drive, Pasadena, CA 91109, USA}

\author{L. Pueyo}
\affil{Space Telescope Science Institute, 3700 San Martin Drive, Baltimore, MD 21218, USA}


\altaffiltext{1}{Jet Propulsion Laboratory, California Institute of Technology, 4800 Oak Grove Drive, Pasadena, CA 91109, USA}



\begin{abstract}
In this paper, we review the impact of small sample statistics on detection thresholds and corresponding confidence levels (CLs) in high contrast imaging at small angles. When looking close to the star, the number of resolution elements decreases rapidly towards small angles. This reduction of the number of degrees of freedom dramatically affects CLs and false alarm probabilities. Naively using the same ideal hypothesis and methods as for larger separations, which are well understood and commonly assume Gaussian noise, can yield up to one order of magnitude error in contrast estimations at fixed CL. The statistical penalty exponentially increases towards very small inner working angles. Even at 5-10 resolution elements from the star, false alarm probabilities can be significantly higher than expected. Here we present a rigorous statistical analysis which ensures robustness of the CL, but also imposes a substantial limitation on corresponding achievable detection limits (thus contrast) at small angles. This unavoidable fundamental statistical effect has a significant impact on current coronagraphic and future high contrast imagers. Finally, the paper concludes with practical recommendations to account for small number statistics when computing the sensitivity to companions at small angles and when exploiting the results of direct imaging planet surveys.
\end{abstract}





\section{Introduction}

Small inner working angle (IWA) coronagraphs are often presented as necessary to take full advantage of existing or planned high contrast imaging instruments, or to efficiently cope with the limited size of space-based telescopes \citep{Roddier1997, Rouan2000, Guyon2003, Mawet2005, Serabyn2010}. In theory, a few coronagraph solutions enable imaging down to the diffraction limit of the telescope (IWA$=1\lambda/D$, i.e.~one resolution element, with $\lambda$ and $D$, the wavelength and telescope diameter, respectively) with sufficient throughput ($\sim50\%$, see \citet{Guyon2006}, or \citet{Mawet2012} for a more recent survey of small angle coronagraphic techniques). However, in order to reach this parameter space, the instrument hosting the coronagraph has to provide exquisite control over low-order aberrations, pointing jitter being the first order perturber, and most difficult to control. This stability requirement puts additional constraints on the instrument and facility, requiring dedicated low-order wavefront/pointing sensors and corresponding correcting elements (mainly tip-tilt and/or deformable mirrors), which often have to be pushed to their hardware and software limits. For a comprehensive review of low-order wavefront sensor solutions chosen by second-generation adaptive optics high contrast imagers, such as GPI \citep{Macintosh2014}, SPHERE \citep{Kasper2012}, SCExAO \citep{Martinache2012} and P1640 \citep{Oppenheimer2013}, among others, see \citet{Mawet2012} for instance. Current first-generation high-contrast imaging instruments are also encroaching on the small angle regime with, e.g., L'-band saturated imaging \citep{Rameau2013}, the Apodizing Phase Plate \citep[APP, see, e.g.,][]{Quanz2010,Kenworthy2010,Kenworthy2013}, the Vector Vortex Coronagraph \citep[VVC, see, e.g.,][]{Serabyn2010, Mawet2011b,Mawet2013, Absil2013}, or the four-quadrant phase-mask coronagraph \citep[FQPM,][]{Riaud2006,Serabyn2009,Boccaletti2012}. 

\subsection{Past work on speckle statistics}
\label{sec:past work}
Statistical tools to assess the significance of a point source detection at large angles are most often based on the assumption that the underlying noise is Gaussian. However, it was noticed a decade ago that speckle noise in raw high contrast images is never Gaussian \citep{Perrin2003,Aime2004,Bloemhof2004,Fitzgerald2006,Soummer2007,Hinkley2007,Marois2008}. The main conclusion of this series of pioneering papers is that the probability density function (PDF) of speckles in raw images does not follow a well-behaved normal (i.e., Gaussian) distribution, but is better described by a modified Rician (MR) distribution, which is a function of the local time-averaged static point-spread function (PSF) intensity $I_c$ and random speckle noise intensities $I_s$:

\begin{equation}\label{eq1}
p_{MR}(I, I_c,I_s)=\frac{1}{I_s}\ \exp{\left( -\frac{I+I_c}{I_s} \right)} \ I_o\left( \frac{2\sqrt{I I_c}}{I_s} \right) 
\end{equation}

where $I_0$ is the modified Bessel function of the first kind, and where the mean and variance of $I$ are $\mu_I=I_c+I_s$, and $\sigma_I^2=I^2_s+2*I_c I_s$, respectively \citep{Soummer2007}.
\begin{figure}[!t]
\centerline{\includegraphics[width=9cm]{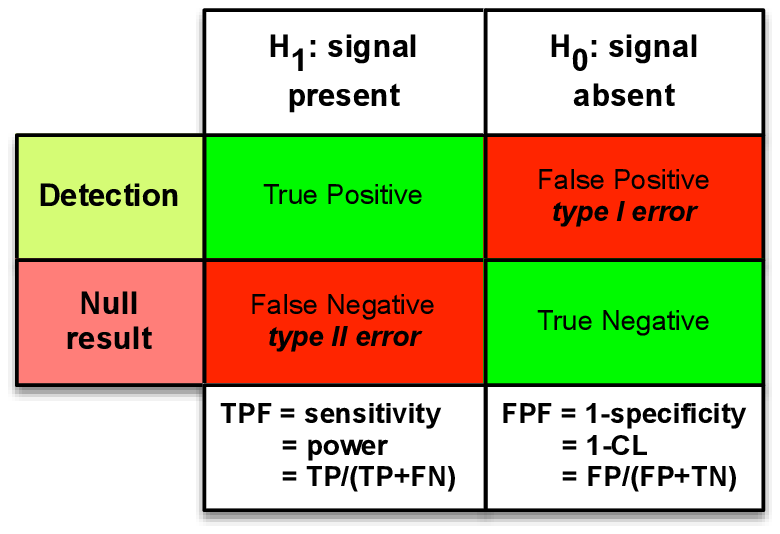}}
\caption{Signal detection theory (SDT) contingency table, or ``confusion matrix''. TP: true positive. FP: false positive. FN: false negative. TN: true negative. TPF: true positive fraction. FPF: false positive fraction. \label{ffig1b}}
\end{figure}

The bulk of past studies related to speckle statistics focused on the temporal aspects of speckle noise in the presence of atmospheric turbulence, corrected or not by adaptive optics systems. In the virtual case of an instrument with perfect optics on a ground-based telescope, the practical impact of the temporal MR PDF of speckles would only have a minor effect on detection limits by virtue of the central limit theorem (CLT). Indeed, if a large number of independent and identically distributed (i.i.d.) speckles are co-added at a specific location in the image, then the sample means will be normally distributed \citep{Marois2008}. In other words, the speckle sampling distribution will be Gaussian.

Unfortunately, optics are never perfect nor hold their imperfect shape constant over time, and so they induce slowly varying wavefront errors, creating quasi-static speckles. \citet{Marois2008} used a heuristic argument to show that quasi-static speckle noise inside annuli centered on the PSF core would follow the MR PDF Eq.~\ref{eq1}, because it is basically produced with the same value of $I_c$ (the unaberrated PSF). The typical lifetime of quasi-static speckles has been found to range from several minutes to hours \citep{Hinkley2007}. $I_{s}$ has thus a complex spatio-temporal dependence $I_{s}(\theta,t)$. Slowly varying wavefront errors disturb the spatio-temporal autocorrelation of the PSF accordingly, and thus its temporal and spatial statistical properties: the samples of resolution elements used to compute noise properties (and thus contrast, see Sect.~\ref{sec:contrastdef}) are no longer independent and identically distributed.

\citet{Marois2008} showed that the net effect of the MR nature of quasi-static speckle noise is that the confidence level (CL) at a fixed detection threshold $\tau$ significantly deteriorates. Subsequently, in order to preserve CLs, or equivalently control the likelihood of type I errors (false alarm probability, or false positive fraction, FPF, see Fig.~\ref{ffig1b}), the detection thresholds (and thus contrast, see Sect.~\ref{sec:contrastdef}) need to be increased significantly, e.g.~up to a factor 4 \citep{Marois2008}. 

\begin{figure}[!t]
\centerline{\includegraphics[width=9cm]{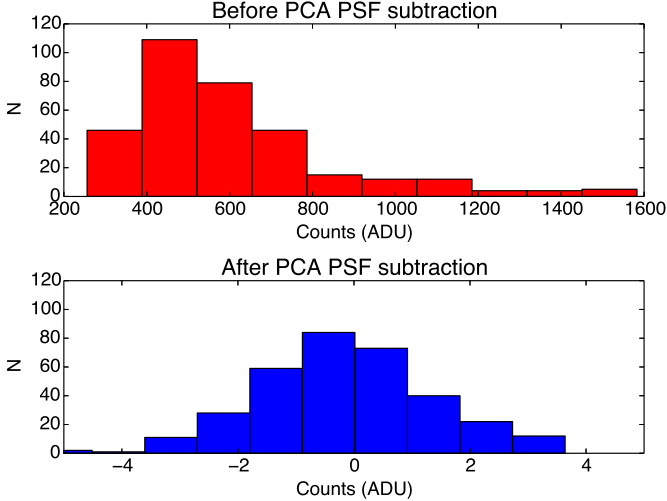}}
\caption{Histograms of pixel values before (up, red) and after (bottom, blue) PCA speckle subtraction and frame co-adding, for the $\beta$ Pictoris data set used as an example in this paper. The statistics of the residual noise between $r=1\lambda/D$ and $r=3\lambda/D$ goes from MR to quasi-Gaussian in this particular example. Indeed, the Shapiro-Wilk test for normality \citep{Shapiro1965} goes from a p-value of virtually 0 to $\sim 0.2$ after speckle subtraction \citep{Absil2013}. \label{ffig1c}}
\end{figure}

Fortunately, observing strategies such as angular differential imaging \citep[ADI,][]{Marois2006}, and data reduction techniques such as the locally optimized combination of images \citep[LOCI,][]{Lafreniere2007, Marois2008} or principal component analysis \citep[PCA,][]{Soummer2012,Meshkat2013} routinely demonstrate their ``whitening'' capability, i.e.~the efficient removal of the correlated component of the noise (see Fig.~\ref{ffig1c}). Whitening yields independent Gaussian noise samples (i.i.d.) through complementary mechanisms. First, once the correlated component has been removed (even partially), other noise sources start to dominate. The latter (background, photon Poisson noise, readout or dark current) are independent noise processes and thus Gaussian by virtue of the CLT. Second, ADI and other differential imaging techniques enhance the efficiency of the first mechanism (if the removal is only partial) by introducing geometrical diversity (field rotation in the case of ADI), further consolidating the independence of noise samples.

\subsection{Posing the problem}
\label{sec:posing}
The present paper starts with the assumption of i.i.d.~noise samples with Gaussian distribution (the non-i.i.d/Gaussian case is examined later) and addresses a different problem: the statistical significance of detections and non detections in the presence of small number statistics, i.e.~when few realizations of spatial speckles vs azimuth are present, which is the case at small angles. In this work, we follow current mainstream practices in the field of high contrast imaging using ground-based adaptive optics facilities, and build up on the past work on speckle statistics presented in Sect.~\ref{sec:past work}. We do not address the choice of statistical paradigm between the frequentist and bayesian approaches (see \citet{Johnson2013} for an interesting review of the subject). Instead, we assume that any prior information available (e.g.~other images produced by differential techniques, reconstructed images from telemetric data, etc.) is used by the data reduction algorithm (e.g.~LOCI or PCA) to obtain data products as whitened as possible given the priors. The statistical analysis is then performed on the whitened products using a classical, frequentist approach.

\begin{figure}[!t]
\includegraphics[width=8.5cm]{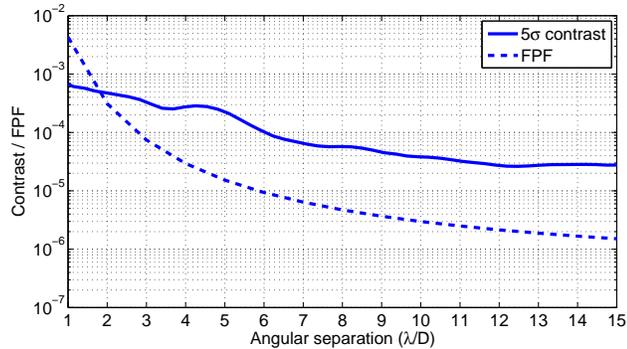}
\includegraphics[width=8.5cm]{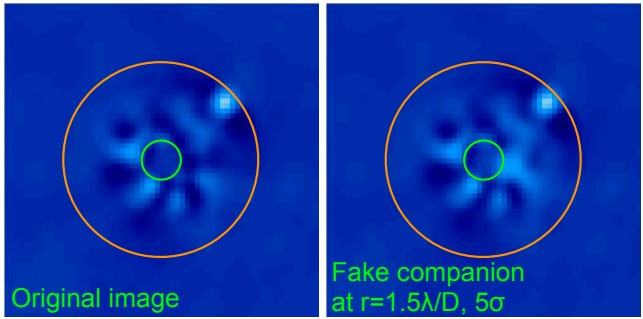}
\caption{$\beta$ Pictoris contrast curve (top image, continuous curve) and image (bottom left, North is not up) taken with NACO in the L-band \citep{Absil2013}, both corrected for the ADI-PCA data reduction throughput. The small green circle is of radius $r=1\lambda/D$, while the big orange one is of radius $r=5\lambda/D$. A fake planet was injected at $r=1.5\lambda/D$ (to the right of the green circle) at the $5\sigma$ throughput-corrected contrast level as presented in \citet{Absil2013}. This $5\sigma$ fake companion is supposedly yielding a solid detection, rejecting the null hypothesis at the $1-3\times 10^{-7}$ CL, assuming normally distributed noise. This is clearly not the case here because of the effect of small sample statistics at small angles. The false positive fraction curve (FPF, dashed line) traces the increase of false alarm probability (or equivalently, the decrease of CL) towards small angles. Note that the scale of the y axis is unique, the contrast and FPF curves being dimensionless. Both quantities are related but have different meanings (see text for details).
\label{ffig2b}}
\end{figure}

Fig.~\ref{ffig2b} presents an excellent illustration of the simple problem at hand. It shows the $\beta$ Pictoris contrast curve and image obtained with NACO in the L-band \citep{Absil2013}, both corrected for the ADI-PCA data reduction throughput. A fake planet was injected at $r=1.5\lambda/D$ at $5\sigma$, where $\sigma$ here is the throughput-corrected contrast level. Using current mainstream contrast definitions assuming well-behaved Gaussian noise (see Sect.~\ref{sec:contrastdef}), this $5\sigma$ fake companion should yield a very reliable detection, i.e.~allowing us to reject the null hypothesis (non-detection) with a $0.99999971 = 1-3\times 10^{-7}$ CL. However, the $5\sigma$ fake companion at $r=1.5\lambda/D$ is barely visible, even when comparing the left and right images side by side. This surprising loss of apparent contrast is primarily due to the limited number of samples in the annulus at $r=1.5\lambda/D$. The present paper aims at quantifying this effect within a rigorous statistical framework, yielding, for instance, the false positive fraction (FPF) dashed curve of Fig.~\ref{ffig2b}. We will present how to rigorously compute the FPF (or equivalently the CL) as a function of angular separation, and show why the FPF (resp.~CL) increases (resp.~decreases) towards small angles. 

The example presented in Fig.~\ref{ffig2b} is the main motivation behind this paper: it is clear that one cannot simply use conventional assumptions and methods used at larger separations anymore. The paper is organized as follows: Sect.~\ref{sec:contrastdef} redefines the notion of contrast and puts it in a rigorous signal detection theory (SDT) statistical framework; Sect.~\ref{sec:smallsample} states the problem of small number statistics in high contrast imaging at small angles; Sect.~\ref{sec:student} is the core of the paper, presenting the Student t-test and corresponding distribution, demonstrating its perfect match to the problem at hands (we also redefine the signal-to-noise ratio) thanks to Monte-Carlo numerical simulations; Sect.~\ref{sec:consequences} follows with a thorough discussion of the consequences and mitigation strategies of the small sample fundamental limitation, and Sect.~\ref{sec:conclusions} presents the conclusions.

\section{Contrast definitions}
\label{sec:contrastdef}

To assess the impact of small number statistics on noise estimation at small angles, and its impact on contrast, we first need a good definition of this metric. On the one hand, contrast can be quantified as the residual intensity $x$, measured either on the attenuated stellar peak (peak-to-peak attenuation), or averaged (mean or median) over different areas of the image, and normalized by the stellar peak intensity. Alternatively, contrast can also be quantified by the ``noise'' measured as the standard deviation $s$ of pixels or resolution elements $\lambda/D$ in a given region of the image, depending on practices and whether the total noise is dominated by the various possible background noise sources, photon or speckle noises. These possible noise measurements are also normalized to the stellar peak intensity to yield relative contrast values. While all these possible definitions can be useful in different contexts (e.g.~technical comparison for the mean intensity), the only relevant metric is however the one that can directly be translated into scientific terms, i.e.~detection limits for putative point sources (or in some cases extended objects) as a function of location relative to the central star. 

Most low-mass companions or exoplanet high contrast imaging studies and surveys have now adopted a $\tau=5\sigma$ detection threshold, which for Gaussian noise is associated with a $\sim 3 \times 10^{-7}$ FPF, or $\sim 1-3 \times 10^{-7}$ specificity (= CL). Following the work of \citet{Marois2008}, it is informally accepted by the high contrast community that this $5\sigma$ level can underestimate the FPF (or overestimate the CL), but it is still used as an easy metric that can be directly compared to other systems. However, one corollary of the present work is that ``all $5\sigma$ contrasts are not equivalent'' in terms of FPF (or CL), which carries the risk of strongly biasing potential comparisons.

\subsection{Signal detection theory}

Referring to the SDT, the detection problem consists in making an informed decision between two hypotheses, $H_0$, signal absent, and $H_1$, signal present (see Fig.~\ref{ffig2b}). The application of hypothesis testing for the binary classification problem of exoplanet imaging was discussed in details by \citet{Kasdin2006}, using matched filtering and Bayesian techniques, but this study focussed on background and photon noise only without any considerations for speckle noise or sample sizes. 

Because most exoplanet hunters want to minimize the risk of announcing false detections or waste precious telescope time following up artifacts, high contrast imaging has mostly been concerned (so far) with choosing a detection threshold $\tau$, defining the contrast which minimizes the FPF, defined as  

\begin{equation}\label{eq2}
FPF=\frac{FP}{TN+FP}=\int_{\tau}^{+\infty} pr(x | H_0) dx
\end{equation}

where $x$ is the intensity of the residual speckles, and $pr(x | H_0)$, the probability density function of $x$ under the null hypothesis $H_0$. FP is the number of false positives and TN, the number of true negatives. Under $H_0$, the confidence level $CL=1-FPF$ is called the ``specificity'' in rigorous statistical terms. However, exoplanet hunters who want to optimize their survey, and derive meaningful conclusions about null results, also wish to maximize the so-called ``True Positive Fraction'' (TPF), or in statistical terms the ``sensitivity'' (some authors refer to ``completeness'', see, e.g.~Wahhaj et al.~2013\nocite{Wahhaj2013}), which is defined as

\begin{equation}\label{eq3}
TPF=\frac{TP}{TP+FN}=\int_{\tau}^{+\infty} pr(x | H_1) dx
\end{equation}

with $pr(x | H_1)$, the probability density function of $x$ under the hypothesis $H_1$, and where TP is the number of true positives and FN, the number of false negatives. For instance, a $95\%$  sensitivity (or completeness) for a given signal $\mu_c$, and detection threshold $\tau$ means that $95\%$ of the objects at the intensity level $\mu_c$ will statistically be recovered from the data (see Sect.~\ref{subsubsec:survey}). Ultimately, the goal of high contrast imaging, as a signal detection application, is to maximize the TPF while minimizing the FPF.  Optimizing detection thus consists in maximizing the so-called AUC, i.e.~the area under the ``Receiver Operating Characteristics'' (ROC) curve. The ROC curve plots the TPF as a function of the FPF. The optimal linear observer, or discriminant, maximizing the AUC is called the Hotelling observer, and can be regarded as a generalization of the familiar prewhitening matched filter (see, for instance \citet{Caucci2007}, or \citet{Lawson2012} for a review). 

\begin{figure}[!t]
\centerline{\includegraphics[width=8.5cm]{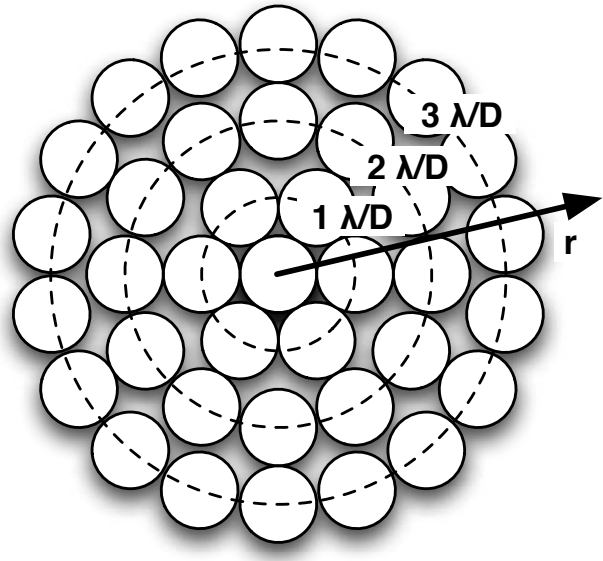}}
\caption{The number of resolution elements at a given radius $r$, is $2\pi r$ (here shown for $r$ ranging from 1 to 3 $\lambda/D$). At close separation, the speckle PDF nature is likely varying drastically as a function of $r$, because of the well-known sensitivity of the PSF to low-order aberrations, especially after a coronagraph. \label{ffig3}}
\end{figure}

\subsection{Small sample statistics}
\label{sec:smallsample}

In the close separation regime (down to the diffraction limit at $1\lambda/D$), speckle noise dominates at all contrast levels, even after being controlled or nulled by active speckle correction \citep{Malbet1995,Borde2006,Giveon2007} and/or a dedicated low-order wavefront sensor (see, e.g., Guyon et al.~2009\nocite{Guyon2009}). In the case of very high contrast images ($10^{9}:1$ and higher), other sources of noise such as photon Poisson noise, readout or dark current might become dominant, especially at larger separations (see, e.g., \citet{Brown2005}, and \citet{Kasdin2006} for thorough treatments of the uniform background case). At small separations, these factors are presumably less important compared to the speckle variability induced by residual low-order aberrations. The detailed error budget largely depends on the hardware available though, and must therefore be studied on a case-by-case basis, which is beyond the scope of this paper.

Quasi-static speckles at a given radius $r$ are all drawn from the same parent population of mean $\mu$ and standard deviation $\sigma$ \citep{Marois2008}. Assuming the detection is performed on individual resolution elements $\lambda/D$, we must treat speckle noise on this characteristic spatial scale as well. We also note that the size of residual speckles is always $\sim \lambda/D$, even after coherent (interference) or incoherent (intensity image) linear combinations. At the radius $r$ (in resolution element units $\lambda/D$), there are $2\pi r$ resolution elements and thus possible non-overlapping speckles, i.e.~about 6 at $1\lambda/D$, 12 at $2\lambda/D$, 18 at $3\lambda/D$, and 100 at  $16\lambda/D$ (see Fig.~\ref{ffig3}). The empirical estimators of the mean and standard deviation, $\bar{x}$ and $s$, are thus calculated from a sample with a limited number of elements (DOF) shrinking with $r$. For samples containing less than $\sim100$ elements (this number is somewhat arbitrary and varies according to practices and applications), we are in the regime of small sample statistics, which significantly affects the calculation of Eq.~\ref{eq2} and Eq.~\ref{eq3}. In this paper, we thus seek to quantify the effect of small sample statistics on the computation of the $pr(x | H_0)$ (and $pr(x | H_1)$), and its impact on the choice of the detection threshold $\tau$, and thus contrast.

In the following, as already discussed, we assume that images have been post-processed by one of the methods presented in Sect.~\ref{sec:past work}. These techniques have been empirically shown to be the most efficient and practical way to use prior information in order to whiten the data. Our working hypothesis in the following is thus that of i.i.d.~samples, so we can focus primarily on the effect of small sample sizes. In Sect.~\ref{sec:montecarlo}, we nevertheless use Monte-Carlo numerical simulations to explore and discuss the consequences of non-i.i.d.~noise (MR distribution) and small sample sizes altogether.

\begin{figure}[!t]
\centerline{\includegraphics[width=8.5cm]{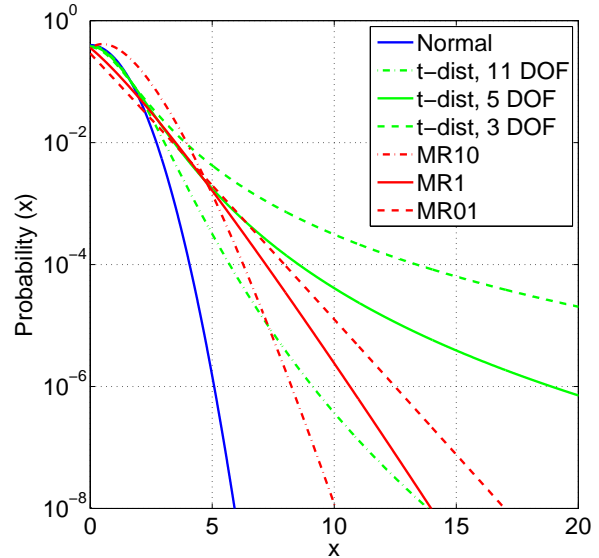}}
\caption{Student's t-distribution PDF (DOF=11,5,3) compared to the normal Gaussian distribution and a few representative MR distributions (MR10: $I_c=10\, I_s$, MR1: $I_c=I_s$, MR01: $I_c=0.1\, I_s$). It illustrates the PDF tail broadening as the number of DOF (sample size minus 1) decreases. Note that no specific normalization was applied to these PDF.\label{ffig4}}
\end{figure}

\section{Student's t-tests}
\label{sec:student}
 
The t-statistics was introduced in 1908 by William S.~Gosset, a chemist working for the Guinness brewery \citep{Student1908}. William S.~Gosset was concerned about comparing different batches of the stout, and developed the t-test, and the t-distribution for that purpose. However, his company forbade him from publishing his findings, so Gosset published his mathematical work under the pseudonym ``Student''.

\subsection{One-sample t-test}

In essence, the one-sample t-test enables us to test whether the mean of a normal parent population has a specific value $\mu$ under a null hypothesis. Gosset showed that the quantity $(\bar{x} - \mu) / (s / \sqrt{n})$, where $\bar{x}$ and $s$ are the empirical mean and standard deviation respectively, and $n$ is the sample size, follows a distribution that he called the ``Student distribution'', or ``t-distribution'', with $n-1$ DOF:

\begin{equation}\label{eq4}
p_{t}(x,\nu)=\frac{\Gamma \left( \frac{\nu+1}{2} \right)} {\sqrt{\nu \pi}\Gamma \left( \frac{\nu}{2} \right) } \left( 1+ \frac{x^2}{\nu} \right)^{-\frac{\nu+1}{2}},
\end{equation}

where $\Gamma$ is the Gamma function, and where the parameter $\nu$ is the number of DOF (here $\nu=n-1$). The one-sample t-test allows accepting or rejecting the null hypothesis once a CL has been set. As a corollary, if one accepts the null hypothesis, a confidence interval on the mean of the parent population can be derived: $\mu \in [\bar{x} - p_t s/\sqrt{n}; \bar{x} + p_t s/\sqrt{n}]$.

The t-distribution $p_t$ is symmetric and bell-shaped, like the normal distribution, but has broader tails, meaning that it is more prone to producing values that fall far from its mean. When $\nu$ is large, Student's t-distribution converges towards the normal distribution (see Fig.~\ref{ffig4}). The t-test is said to be robust to moderate violations of the normality assumption for the underlying population \citep{Student1908, Lange1989}. Indeed, the parent population does not need to be normally distributed, but the population of empirical sample means $\bar{x}$ (i.e.~the sampling distribution), is assumed to be normal by the CLT, therefore valid for reasonably large samples. This particularly interesting property will be put to the test in Sect.~\ref{sec:montecarlo}. 

\begin{figure*}[!t]
\begin{center}
\includegraphics[width=8.5cm]{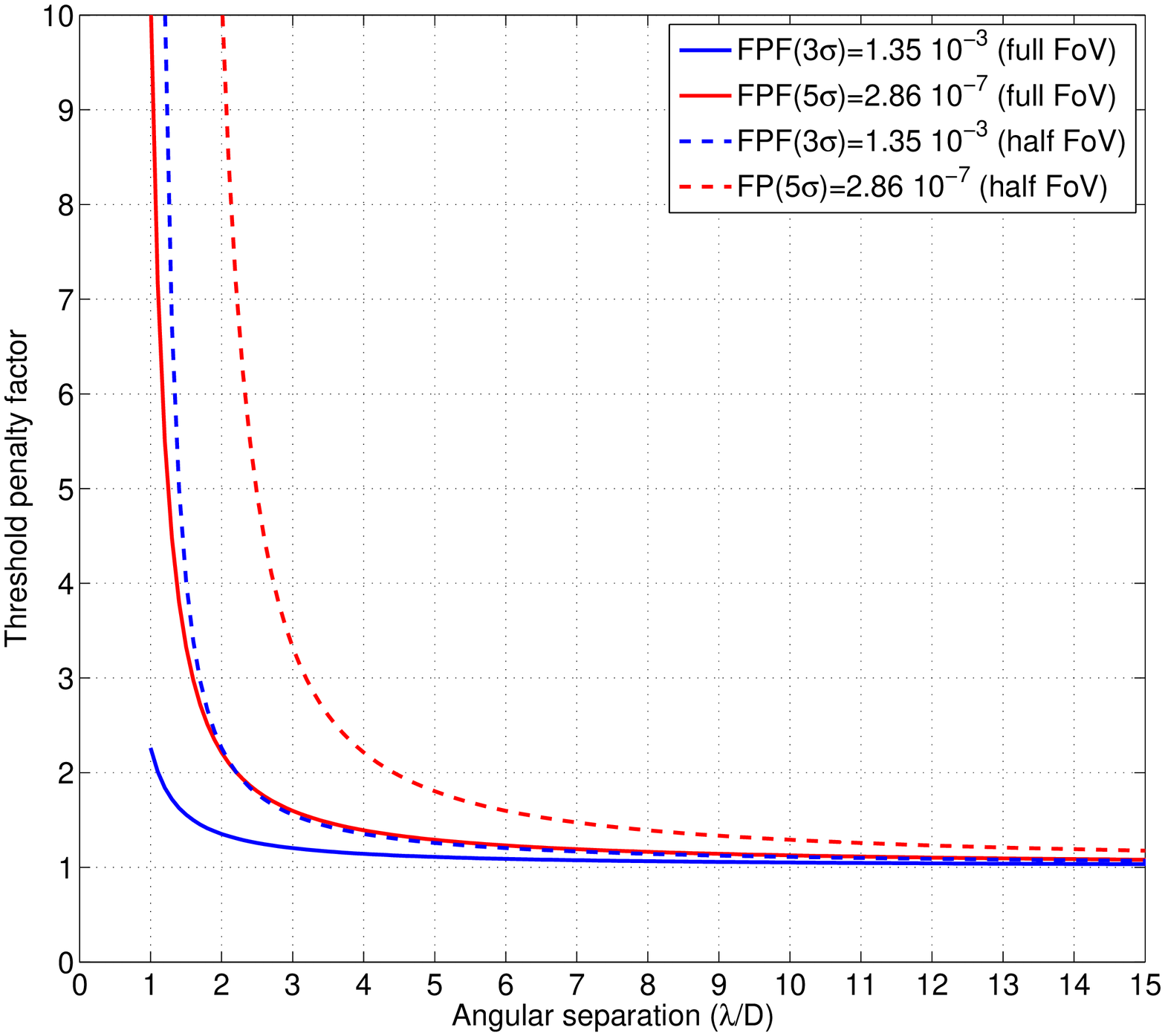}
\hspace{0.5cm}
\includegraphics[width=8.5cm]{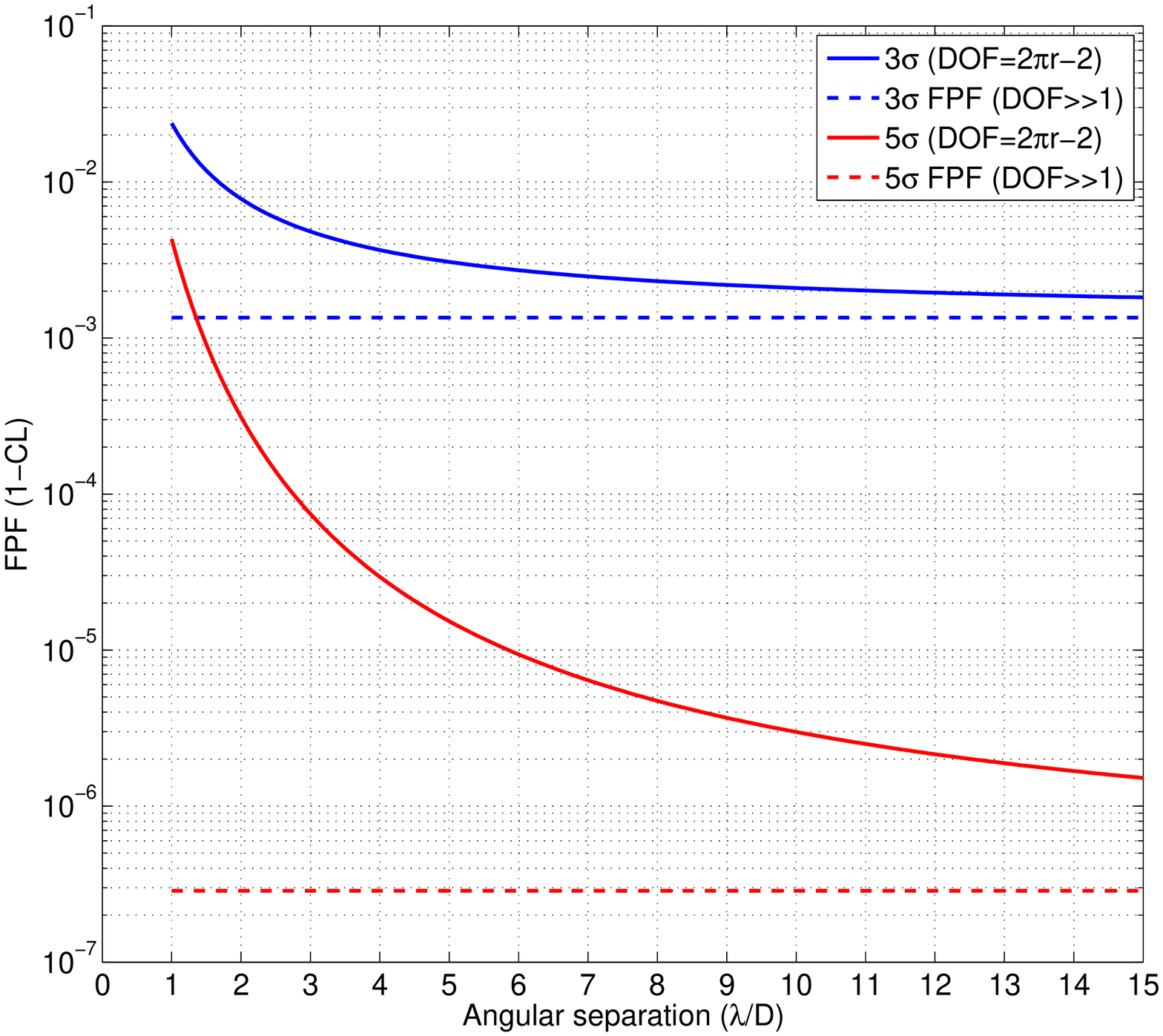}
\caption{Left: $1-3\times 10^{-7}$ CL detection threshold for small sample statistics as a function of angular separation (in $\lambda/D$), divided by 5 (yielding the correction to be applied to the usual 5$\sigma$ Gaussian threshold). The two curves correspond to the full FoV and half FoV cases. Right: FPF ($1-$CL) for a $5\sigma$ detection threshold as a function of angular separation, using the t-distribution. \label{ffig5}}
\end{center}
\end{figure*}

\subsection{Two-sample t-test}

The detection process can be viewed as a test comparing one resolution element at a time (sample \#1) against all the remaining $n-1$ ones  (sample \#2) at the same radius $r$ (again, $r$ is measured in resolution element units $\lambda/D$). Under the null hypothesis, one can verify that these two samples are indeed drawn from a common parent population of unknown $(\mu,\sigma^2)$ by comparing their empirical sample means $\bar{x}_{1}$ and $\bar{x}_{2}$. Verifying the null hypothesis that two sample means are equal is the essence of Gosset's ``two-sample t-test''.

So far, and except for the work in \citet{Marois2008}, FPF (and thus corresponding contrast) calculations have always assumed normally distributed speckle statistics and large sample sizes, and therefore a virtually perfect knowledge of the underlying parent population of speckles $(\mu,\sigma^2)$. Within this oversimplified framework, a speckle population of mean $\mu$, and standard deviation $\sigma$, produces the corresponding FPF simply given by

\begin{equation}\label{eq5}
FPF=\int_{\tau}^{+\infty} pr(x | H_0) dx=\int_{\tau}^{+\infty} \mathcal{N}(\mu,\sigma^2) dx
\end{equation}

where 
\begin{equation}\label{eq5b}
\mathcal{N}(\mu,\sigma^2)=\frac{1}{\sqrt{2\pi}\sigma} e^{-\frac{1}{2}\left(\frac{x-\mu}{\sigma}\right)^2}
\end{equation}

For instance, as mentioned earlier, with $\tau=5\sigma$, we have the now classically adopted false alarm probability of $\sim 3\times 10^{-7}$. With $\tau=3\sigma$, we have a false alarm probability of $\sim 1.35\times 10^{-3}$. 

Now, when the parent population characteristics $(\mu,\sigma^2)$ are unknown and sample sizes small, one has to use the corresponding empirical estimators ($\bar{x}$ and $s$), and the t-test for unequal sample sizes, equal variances (thus assuming homoscedasticity, i.e.~homogeneity of variance, see below)

\begin{equation}\label{eq6}
p_t(x,\nu) \equiv \frac{\bar{x}_{1}-\bar{x}_{2}}{s_{1,2}\sqrt{\frac{1}{n_1}+\frac{1}{n_2}}},
\end{equation}

where $\bar{x}_{1}$ is the intensity of the single test resolution element ($n_1=1$), and $\bar{x}_{2}$ is the average intensity over the remaining $n_2=n-1$ resolution elements in a $1\lambda/D$-wide annulus at the radius $r$, and where

\begin{equation}\label{eq7}
s_{1,2}=\sqrt{\frac{(n_1-1)\sum\limits_{i=1}^{n_1} \frac{(x_i-\bar{x}_{1})^2}{n_1-1}+(n_2-1)\sum\limits_{i=1}^{n_2} \frac{(x_i-\bar{x}_{2})^2}{n_2-1}}{n_1+n_2-2}}
\end{equation}

The pooled standard deviation $s_{1,2}=s_{2}$ for $n_1=1$. $s_2$ is the empirical standard deviation computed over the remaining $n_2=n-1$ resolution elements at radius $r$. Our initial hypothesis of homoscedasticity is warranted twice. First, under the null hypothesis, we want to verify that resolution element samples at a given radius $r$ (measured in $\lambda/D$ units) are drawn from a parent population of speckles, with an unknown but common variance $\sigma^2(r)$. To comply with this statement, any detection should of course be excluded from the sample of remaining $n-1$ resolution elements to prevent biases. Second, the presence of a bona fide companion at the location of the test resolution element $x_1$ will only change the mean but not the variance of the underlying population.

One might also question the significance of the two-sample t-test, when one of the test samples only has a single element ($n_1=1$). However, the numerical simulations presented in Sect.~\ref{sec:montecarlo} empirically demonstrate its applicability in such a particular configuration. Note that resolution elements are treated independently of any pixel sampling considerations, which in practice is equivalent to binning the data by the pixel sampling before applying the t-test. Substituting Eq.~\ref{eq7} into Eq.~\ref{eq6}, we have the formal t-test for high contrast imaging at small angles

\begin{equation}\label{eq8}
p_t(x,n_2-1) \equiv \frac{\bar{x}_{1}-\bar{x}_{2}}{s_{2}\sqrt{1+\frac{1}{n_2}}},
\end{equation}

yielding the FPF or false alarm probability, now depending on $\nu=2\pi r-2$ DOF (indeed, $n_2-1=n-2$, with $n=\text{round}(2\pi r)$, and r measured in $\lambda/D$ units), 
\begin{equation}\label{eq9}
FPF=\int_{\tau}^{+\infty} pr(x | H_0) dx=\int_{\tau}^{+\infty} p_t(x,n_2-1) dx
\end{equation}

We note the similarity of Eq.~\ref{eq8} to the standard signal-to-noise ratio (SNR) definition in high contrast imaging (see, e.g., Rameau et al.~2013\nocite{Rameau2013}), except for the $\sqrt{1+1/n_2}$ correction factor to the empirical standard deviation $s_2$, and of course the equality to the Student t-distribution with $n_2-1=n-2$ DOF instead of the normal distribution. 

We argue that Eq.~\ref{eq8} is the true definition of SNR, which can be rigorously linked to CLs through $p_t(x,n_2-1)$, and recommend its use from now on. It is also worth emphasizing that Eq.~\ref{eq8} converges to the standard definition of SNR for large sample sizes since the correction factor $\sqrt{1+1/n_2}$ converges to 1 for $n_2>>1$, and the t-distribution converges to the normal distribution for $DOF>>1$. Of course, this convergence does not imply that the underlying noise in the images becomes gaussian.

The effect of small sample statistics, rigorously described by the t-distribution, is to broaden the tails of the effective speckle PDF, raising the fixed-CL detection thresholds, and thus contrast accordingly. It is important to note that, contrary to the MR distribution which describes the true nature of speckle noise, the t-distribution only describes our fundamental incapacity to characterize it, due to the lack of information. This effect can be significant, and yields a factor 10 penalty for the classically calculated\footnote{Note that, in the case of small samples, the standard deviation of the parent population $\sigma$ (the noise) is unknown, so we should use the empirical standard deviation $s$. However, for the sake of simplicity, we will use the conventional notation $\sigma$ in the following when actually referring to the empirical standard deviation $s$.} $5\sigma$ ($\text{FPF}\simeq 3\times 10^{-7}$) contrast limit at $1\lambda/D$, and factor of 2 degradation at $2\lambda/D$ (see Fig.~\ref{ffig5}, left). Penalty factors are significantly reduced if one adopts a less stringent threshold, for instance $3\sigma$ (see Fig.~\ref{ffig5}, left). Note that in some cases, only half of the field of view (FoV) is accessible, as with, e.g., the APP \citep{Quanz2010,Kenworthy2010,Kenworthy2013} or half dark holes \citep{Malbet1995,Borde2006,Giveon2007}, reducing the number of DOF by another factor of two, penalizing contrast thresholds and FPF/CLs even more (see Fig.~\ref{ffig5}, left, dashed curves). 

However, one can argue that a $3\times 10^{-7}$ false alarm probability might not really be required at small IWA. 
Indeed, since the statistical tests and corresponding results discussed here are done resolution element by resolution element, the total number of potential false alarms at a given radius $r$ is proportional to $\text{FPF}(r) \times 2\pi r$. It is therefore interesting to fix the detection threshold to $5\sigma$ or $3\sigma$ (whatever the PDF) and derive how the CL evolves in the small sample statistics case, described by the t-distribution (see Fig.~\ref{ffig5}, right). At $r=1\lambda/D$, a $5\sigma$ detection threshold still yields $\sim 0.004$ false alarm probability, which is fairly close to (but still a factor 3 above) the nominal $3\sigma$ Gaussian false alarm probability.

\begin{figure}[!t]
\centerline{\includegraphics[width=8.5cm]{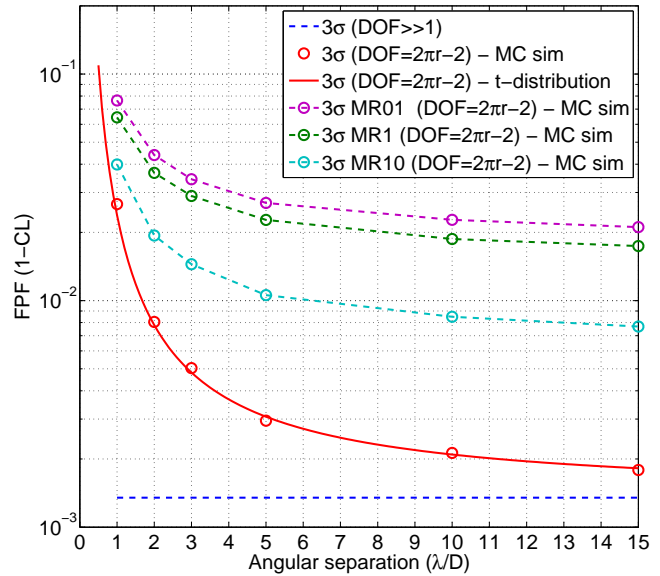}}
\caption{False positive fractions (false alarm probabilities) for the canonical t-distribution and Monte-Carlo numerical simulations, demonstrating the validity of the t-distribution at small angles in the normal case. For MR speckle statistics however, the t-distribution is still underestimating the FPF, despite the very broad tails of its PDF. \label{ffig6}}
\end{figure}

\subsection{Monte-Carlo simulations}
\label{sec:montecarlo}
We proceeded with Monte-Carlo numerical simulations for two reasons:
\begin{enumerate}
\item Verify that the two-sample t-test is indeed valid if one of the two samples only has a single element.
\item Test the robustness of the t-test when the residual speckle noise follows a MR distribution.
\end{enumerate}
We generated various random samples of size $n=2\pi r$ issued from normal and MR PDFs, effectively simulating random speckle samples as a function of the radius $r$ (in $\lambda/D$ units). We then used the two-sample t-test (Eq.~\ref{eq8}) and tested each simulated resolution element $\bar{x}_1$ against the mean $\bar{x}_2$ and standard deviation $s_2$ of the remaining ones (excluding the resolution element under test), repeated this test many times, and counted the number of false positives, i.e.~with 
$$(\bar{x}_1-\bar{x}_2) > 3 \times s_{2} \sqrt{1+\frac{1}{n_2}} $$ We chose to use a ``$\tau = 3\sigma$ threshold'' here to avoid prohibitive computation times.

The results of these simulations are summarized in Fig.~\ref{ffig6}. It shows the perfect agreement between the t-distribution and the measured false alarm probability for a normally distributed parent population of speckles. The simulations thus demonstrate the applicability of Student's two-sample t-test when one of the samples only has one element. For the MR cases however, the t-distribution underestimates the false positive fraction by a significant factor, both at small and large angles. This empirical result, reminiscent of the results presented in \citet{Marois2008} for large samples and using a complementary methodology, is not surprising, as the MR PDF statistically describes the spatio-temporal autocorrelation of the PSF. The PSF autocorrelation invalidates our working hypothesis of i.i.d.~samples, which is also an important pre-condition for the applicability of the Student t-test. 

In other words, the Student t-test might be robust to slightly non-Gaussian underlying population, but this property cannot be verified or used here because the Student t-test is NOT robust to non-i.i.d.~samples. This stems from the fact that the so-called robustness of the  t-test is rooted in the CLT, which requires i.i.d.~samples. Non-Gaussian speckle statistics, hopefully a rare occurrence when the data reduction is performed efficiently, therefore needs its own particular solution: in Sect.~\ref{subsubsec:collection}, we briefly introduce and discuss non-parametric tests such as the Wilcoxon signed-rank and rank-sum tests, also known as the Mann-Whitney U test \citep{Wilcoxon1945}.

\begin{figure}[!t]
\includegraphics[width=8.5cm]{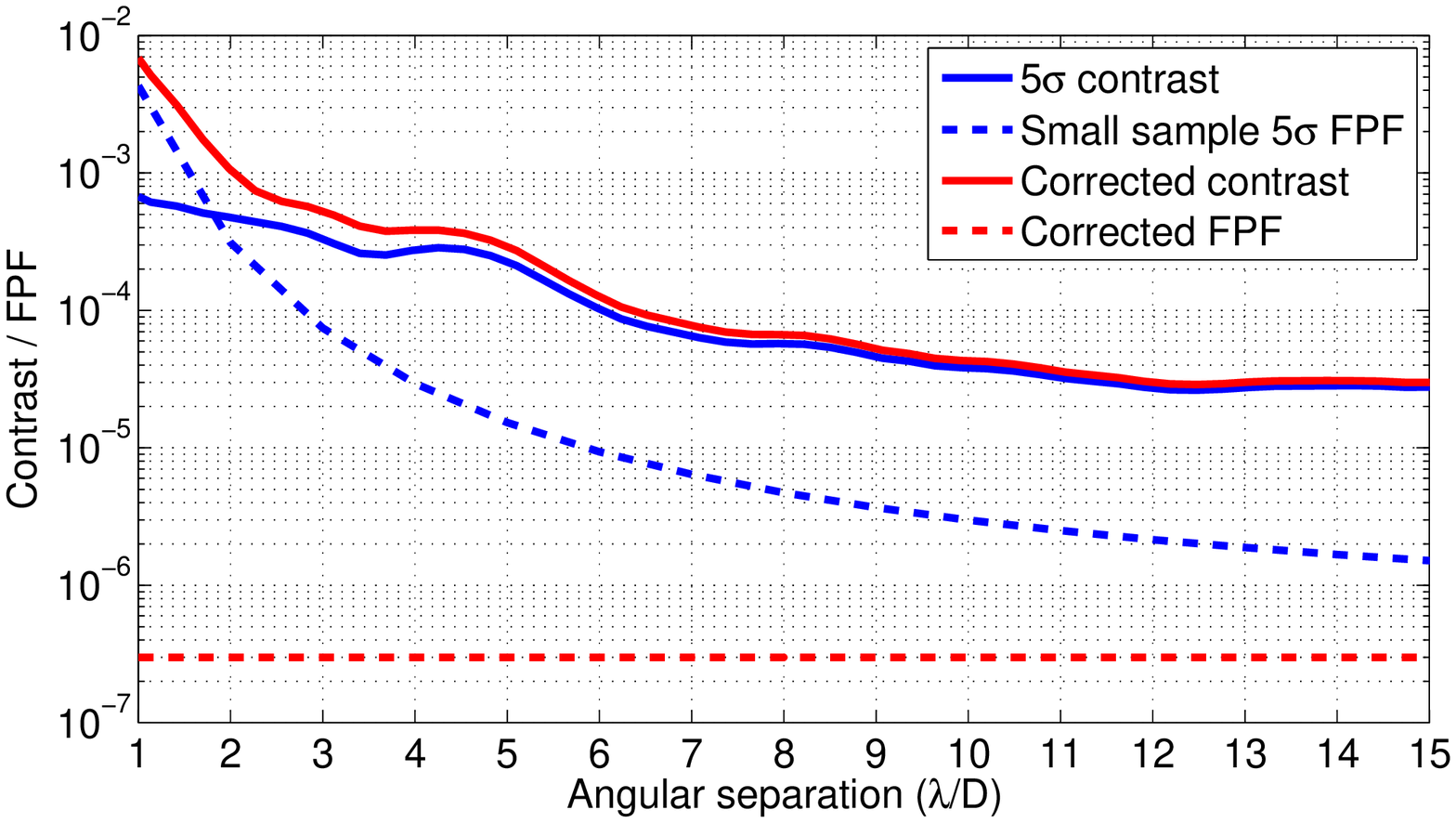}
\includegraphics[width=8.5cm]{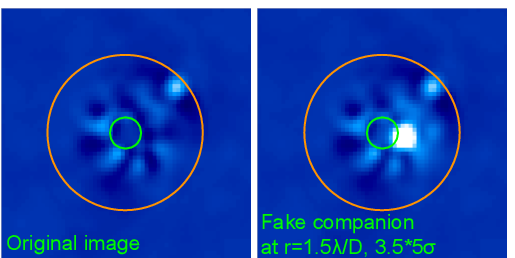}
\caption{Same data as in Fig.~\ref{ffig2b}, now corrected for the effect of small sample statistics. The red contrast curve is showing the true contrast with $3\times 10^{-7}$ FPF (dashed red curve). A penalty factor of 3.5, following Student's t-test, has now been applied to the fake companion in the bottom right image, restoring a $1-3\times 10^{-7}$ CL for rejecting the null hypothesis.
\label{ffig7}}
\end{figure}

\subsection{From t-test to 1D contrast curves}
\label{sec:ttest2contrast}

Assuming i.i.d.~samples, the two-sample t-test allows testing the $H_0$ hypothesis one resolution element at a time in a rigorous statistical framework, accounting for small sample sizes. However, testing resolution elements one by one to generate a 2D contrast map can be tedious and not very relevant in case of non-detection. A common practice in high-contrast imaging is to generate 1D contrast curves, and so here we provide a simple recipe for calculating contrast under the null hypothesis:

\begin{enumerate}
\item Choose a maximum FPF or confidence level CL.
\item Compute the mean $x$ and standard deviation $s$ of resolution elements at radius $r$, along with the number of resolution elements $n=round(2\pi r)$. 
\item From Eq.~\ref{eq9}, solve 
\begin{equation}\label{eq9b}
FPF=\int_{\tau}^{+\infty} pr(x | H_0) dx=\int_{\tau}^{+\infty} p_t(x,n-1) dx 
\end{equation}
for $\tau$.

\item Multiply $\tau$ by $s \sqrt{1+1/n}$, then add $x$ (usually $x\approx 0$).
\end{enumerate}

Solving Eq.~\ref{eq9b} requires a numerical integration, which is available as a standalone routine in most languages (e.g., ``$t\_cvf(CL,n-1)$'' in IDL, but there are similar functions in python/numpy, matlab, R, excel, etc.). 

The only difference between the t-test conducted one resolution element at a time and the proposed 1D contrast curve recipe, is that the latter loses the azimuthal spatial information. Subsequently, the localization of the 2D two-sample t-test is traded off with the gain of an additional DOF ($n-1$ instead of $n-2$), which slightly relax the contrast penalties.

\section{Consequences and mitigation strategies}
\label{sec:consequences}
In this section, we discuss the consequences of small sample statistics on contrast limits for single objects, and surveys. We also provide mitigation strategies to try and overcome the limitations imposed by small samples, and to ensure robustness of contrast estimations.

\subsection{Single object}

When investigating the detection limits for a single object or to decide whether or not a candidate shall be followed up, one is subject to direct hits from the limited number of samples available. In the following, we will distinguish the case where the contrast computation is done on a single image (e.g.~final product of a pipeline), and the case where it is conducted on parts of, or all of the individual frames from the observing sequence.

\subsubsection{Case of one object with a single image available}\label{subsubsec:single}

The case of one object with a single final image available for the detection limit analysis is the worst case scenario since the amount of information is extremely limited. This situation is however unlikely, and would only occur if one does not have access to, or master the inner mechanics of a third-party pipeline. It could also occur in the future for very high contrast imaging coronagraphs on small space telescopes (1-2.4 meters), where contrast levels are so high, and telescopes relatively small, that exposures become long and scarce (they are limited in time due to cosmic rays though). 

In this limiting case, detection limits would directly be affected by the Student t-distribution, with penalty factors (with respect to the normal 5$\sigma$ detection threshold, i.e.~with $3\times 10^{-7}$ FPF) as high as $\sim 10$ at 1$\lambda/D$, but would decrease to $\sim 2$ at 2$\lambda/D$, assuming purely gaussian noise (see Fig.~\ref{ffig5}, left). Fig.~\ref{ffig7} showcases a practical example using the same data as in Fig.~\ref{ffig2b} but now with a contrast curve corrected for the effect of small sample statistics, and a fake companion injected at the level prescribed by the t-distribution in order to preserve confidence levels. The detection is now much clearer than in Fig.~\ref{ffig2b}, confidence levels are restored.

In the eventuality of non-i.i.d.~samples from a non-Gaussian underlying population, we have demonstrated that the significance of the t-test is limited, although it is nevertheless much more conservative than current practices. The Wilcoxon signed-rank and rank-sum tests \citep{Wilcoxon1945} are non-parametric tests that can be used as alternatives to Student's t-tests when populations cannot be assumed to be normally distributed, for dependent (paired) and independent samples, respectively. However, the Wilcoxon rank-sum test\footnote{The Wilcoxon rank-sum test proceeds as follows: arrange the data values from both samples under test in a single ascending list, assign the numbers 1 to $N$ (where $N = n_1+n_2$). These are the ranks of the observations. Let $W_1$ and $W_2$ denote the sum of the ranks for the observation from sample 1 and 2, respectively. The Mann-Whitney statistics for sample 1 and 2 are defined as follows: $U_{1,2}=n_1 n_2+n_{1,2}(n_{1,2}+1)/2-W_{1,2}$, respectively. If there is no difference between the two medians (the null hypothesis), the value of $W_1$ and $W_2$ will be around half the sum of the ranks $(n_{1,2}(1+N))/2$. The statistics $Z=(U_{1,2}-(n_1 n_2)/2) / \sqrt{n_1 n_2 (n_1+n_2+1)/12}$, follows a normal distribution for reasonably large sample sizes. For very small sample sizes, one must refer to tabulated values of the Mann-Whitney statistics $U$.}, for instance, loses significance if one of the two samples has a single element. The relevance of these non-parametric methods and resampling/bootstrapping \citep{Loh2008} in contrast estimations requires more work and will be the subject of future research. For now, when dealing with extremely small sample sizes, one has to verify or assume that the samples are sufficiently i.i.d.~and normally distributed. For that, there are several well-known non-parametric tests one can use to verify a priori that the limited sample at hand came from a normally distributed population. For instance, the frequentist Shapiro-Wilk test \citep{Shapiro1965}, which has been proven to have the best power for a given significance \citep{Razali2012}, was used in \citet{Absil2013}. If there is evidence that the population is non-normal, the only alternative is to gather more data to either further whiten the noise, or increase the sample size to better constrain the PDF altogether.

\subsubsection{Case of one object with a collection of images}
\label{subsubsec:collection}
Usually, the observing sequence of a single object consists of several dozens of images, most of the time combined (averaged) into a single final frame, where small sample statistics effects are substantially affecting detection limits as we just saw. However, it is conceivable that the analysis is conducted on the ensemble of individual frames, increasing the number of DOF accordingly, and therefore alleviating the effect of small sample statistics\footnote{We note here that auxiliary measurements such as telemetric data from wavefront sensors can in principle have the same role as additional frames.}. There are however three important caveats: photon noise, decorrelation timescales for quasi-static speckles (as discussed in the introduction), and human factors in signal detection if target vetting is done visually (as is often the case). 

Let us introduce the debinning factor $\zeta$, which gives the final number of images retained for analysis. We note that in this case, $n_1$ that was equal to 1 in the previous case will now be larger. Indeed we now have $n_1^* = n_1 \zeta$ and $n_2^* = n_2 \zeta$.

\paragraph{Photon noise vs small samples statistics}

\begin{figure}[!t]
\centerline{\includegraphics[width=8.5cm]{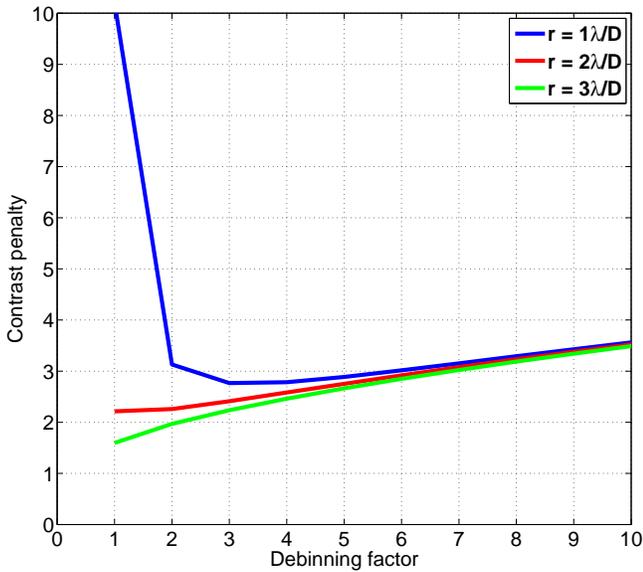}}
\caption{Contrast penalty as a function of debinning factor ($\zeta$), for 3 different radii $r$. The contrast penalty ($\tau s$) combines the effect of the threshold penalty due to small sample statistics $\tau$, and the photon noise-induced increase of $s$ in $\sqrt{\zeta}$. The major caveat of this calculation is that we assumed well behaved decorrelated images, which is dubious especially at small angles. \label{ffig8}}
\end{figure}

If the final combination (averaging) of images is prevented, one has to consider the effect of photon noise (neglected so far) which affects SNR of individual images as $\sqrt{\zeta}$. Figure~\ref{ffig8} conceptually illustrates the trade-off between increasing the sample size and photon noise for various debinning factors (e.g.~a debinning factor of 2 means that the whole data set was binned in two combined frames), assuming decorrelated images. From this ideal case, the trade-off yields a minimum penalty factor at a debinning $\zeta \simeq 3$ for $r=1\lambda/D$. Beyond $r=1\lambda/D$ however, there is no gain brought by debinning because the detrimental effect of photon noise dominates the detrimental effect of small sample statistics.

\paragraph{Residual correlated noise vs small sample statistics}

For a series of exposures taken on a single object, it still may happen that the reduced individual images are not i.i.d.~realizations with a well behaved Gaussian noise. This situation, while unlikely, could occur despite best whitening efforts, especially at small angles. Indeed, while ADI is very efficient at large angles, the limited projected parallactic angle variation at small angles might somewhat impair efficient whitening by the second mechanism mentioned in Sect.~\ref{sec:past work}, leaving the remaining speckles potentially affected by residual correlated noise (MR PDF). 

Any statistical inference based on these non-recombined images therefore requires to take the actual temporal PDF of the residual speckles into account. The PDF of the speckles very close to the center can be very difficult to determine (it depends on the control system, observing conditions and strategy, data reduction technique, etc.). Given the radial dependence of the PSF and ADI (and similar differential imaging methods), the speckle PDF is also function of radius, precluding radial extrapolations. 

As discussed in Sect.~\ref{subsubsec:single}, if the exact distribution cannot be determined, one can use the Shapiro-Wilk test \citep{Shapiro1965} to ensure normality. If this simple test fails, one can eventually consider the non-parametric Wilcoxon rank tests, which becomes relevant again here because $n_1^* = n_1 \zeta>1$, or bootstrapping. A comparison between these non-parametric methods is out of the scope of the present paper, and is deferred to future work.

\paragraph{Visual vetting: the human factor}

Debinning the data can be useful from a statistical point of view between 1 and 2 $\lambda/D$. However, we argue here that if the final vetting is done by a human looking at an image, or a collection of reduced images (without doing further processing), the t-test is very representative of the human perception of signal hidden in small noisy samples. Fig.~\ref{ffig2b} is a representative illustration of this behavior. Moreover, looking at a collection of images where speckles vary (if whitening was well done) or appear static (inefficient whitening) will not help the visual cortex overcome the small sample statistics effects. 

\subsubsection{Summary}

Within our assumption of i.i.d./whitened data products using any prior information available (through, e.g.~ADI), the t-test is the best practical solution proposed here. Debinning should be considered in the 1-to-2 $\lambda/D$ angular separation region to increase the number of DOF, but one quickly has to face photon noise limitations. The consequences of small sample statistics for very small IWA thus range from severe to acceptable, depending on the final application and total information available. If spatio-temporal correlations remain (e.g.~MR) despite efforts to whiten the data products, the Student t-test will not provide correct significance levels. As mentioned above, the non-parametric Wilcoxon rank tests or bootstrapping should be considered instead, but only if there is more than one element in the test sample.

\subsection{Many objects, surveys}

There are several important factors to consider when analyzing many objects as part, for instance, of a survey conducted in order to statistically constraint populations of low-mass stellar companions, brown dwarfs, planets, or even circumstellar disks.

\subsubsection{More information is good}

The amount of information provided by a survey, or analysis of archival data could in principle alleviate small sample statistics effects, by two complementary mechanisms. First, if there is indeed residual correlated noise (super-static speckles), it will be more easily characterized because of the large sample available (the empirical PDF will be better sampled). Second, if correlated noise is still present in the data, it should also be straightforward to remove it with methods such as a PCA of the PSF library provided by the survey targets \citep{Soummer2012}.

\subsubsection{Alternative definition of contrast relevant to surveys}
\label{subsubsec:survey}
Some authors \citep[e.g.,][]{Wahhaj2013} argue that the TPF (completeness) is more relevant than the FPF, especially when using a survey to obtain planet population constraints. Indeed, their argument is, since detected companions are observed a second time to check for common proper motion with the primary, the chance of a repeated false detection is $\sim$ FPF$^2$ (assuming both observations are uncorrelated, which can be the case, see \citet{Milli2014} for instance), and thus the combined FPF is small. In other words, to derive planet population constraints, one should mainly be concerned about the probability of detecting a planetary object with a given mass, and thus contrast at a given separation, which is different than what Eq.~\ref{eq8} defines, i.e.~the contrast under which we accept the null hypothesis with a given CL. The completeness contrast for a desired TPF (e.g.~95\%) and detection threshold $\tau$ is defined as $\mu_c$, obtained from Eq.~\ref{eq3}. So we have,

\begin{equation}\label{eq10}
\int_{\mu_c-\tau}^{+\infty} pr(x | H_1, \mu_c) dx = TPF
\end{equation}

\begin{figure}[!t]
\centerline{\includegraphics[width=8.5cm]{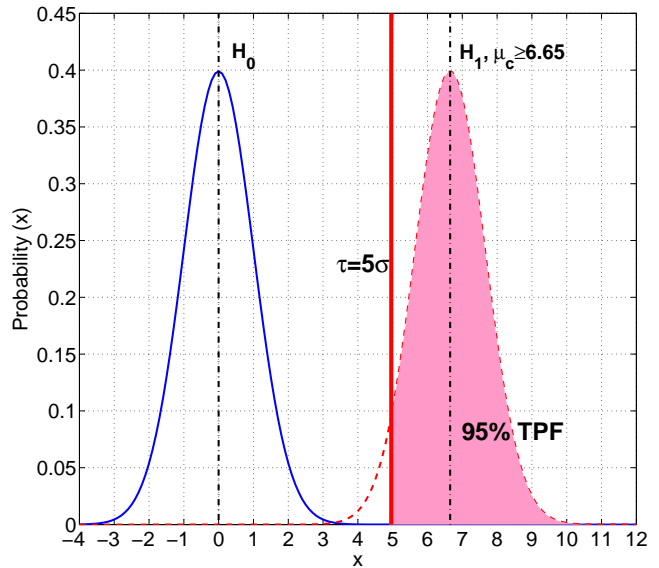}}
\caption{Formal SDT definition of completeness (or sensitivity) for a TPF of $95\%$ with a detection threshold set at $\tau=5\sigma$. The blue curve is the intensity distribution under the signal absent hypothesis $H_0$, and the red curve is the intensity distribution under the signal present hypothesis $H_1$. The $95\%$ completeness (pink area) at a $5\sigma$ threshold is for signals $\mu_c\simeq 5\sigma+1.65\sigma = 6.65\sigma$. \label{ffig9}}
\end{figure}

Fig.~\ref{ffig9} visually illustrates the SDT definition of $95\%$ completeness (or sensitivity) based on Eq.~\ref{eq10}, for a threshold set at $\tau=5\sigma$. According to this definition, the $95\%$ completeness level $\mu_c$ is always $\sim 1.65$ above the threshold $\tau$. Following the SDT formalism, the effect of small sample statistics can then easily be calculated, as one only needs to substitute $pr(x | H_1, \mu_c)$ with the Student t-distribution instead of the normal distribution. The penalties  at small angles are significant but not dramatic (see Table~\ref{table1}).

A low false alarm probability ensures one does not waste time following false detections. At small IWA, near typical target stars in exoplanet surveys, the probability of detecting background stars is very small (this probability can easily be calculated using population models for the galaxy such as TRILEGAL, see, e.g.~Vanhollebeke et al.~2009\nocite{Vanhollebeke2009}). Thus, setting a $3\sigma$ detection threshold at small IWA yields a negligible number of background sources to follow up and only $\sim 2\%$ false detections (and hopefully some real companion detections). Thus wasteful follow-up in this case would be minimal. Note that this argument would only be partially valid for the E-ELT or a space-based coronagraph, where telescope time is very costly, and the competition significant. The completeness contrast (e.g., contrast at which 95\% of real objects are recovered, see Wahhaj et al.~2013\nocite{Wahhaj2013}), should be the main concern when deriving population constraints from a null results survey. Therefore, given the $\mu_c-\tau$ difference, we argue here that a lower threshold level is recommended at small angles, for instance $\tau=3\sigma$, which would yield $95\%$ completeness for sources the $3\sigma+2.1\sigma \simeq 5\sigma$ level, in the worst case, i.e.~the smallest possible angle $1\lambda/D$ (see table~\ref{table1}).

\begin{deluxetable}{llll}
\tabletypesize{\scriptsize}
\tablecaption{The effect of small sample statistics on specificity and sensitivity for high contrast imaging at small angles.\label{table1}} 
\tablehead{
\colhead{Radius ($\lambda/D$)} &\colhead{$\mu_c-\tau$(\tablenotemark{a})} &\colhead{$\tau^{5\sigma}$ (\tablenotemark{b})} &\colhead{$\tau^{3\sigma}$ (\tablenotemark{c})} }
\startdata
1	&2.1		&10		&2.26 \\
2	&1.8		&2.2		&1.35 \\
3	&1.74	&1.6		&1.2 \\
4	&1.71	&1.4		&1.14 \\
5	&1.70	&1.3		&1.1 \\
10	&1.67	&1.12	&1.05 \\
$\infty$	&1.65	&1.0		&1.0 \\

\enddata
\tablenotetext{a}{ To achieve 95\% completeness at $\mu_c$, the threshold must be set this many sigma below $\mu_c$.}
\tablenotetext{b}{Penalty factor for the contrast at $3\times 10^{-7}$ FPF ($5\sigma$ Gaussian) to account for small sample statistics.}
\tablenotetext{c}{Penalty factor for the contrast at $1.35\times 10^{-3}$ FPF ($3\sigma$ Gaussian) to account for small sample statistics.}
\end{deluxetable} 
   
\section{Conclusion}
\label{sec:conclusions}

The penalty factor induced by small sample statistics on contrast computed at very small angles has been presented for the first time. Its impact on the determination of detection limits has been reviewed for several practical cases (single object, one or multiple frames, and surveys). Our recommendation is to use the robust Student t-distribution to mitigate the poor knowledge of speckle statistics at small angles, after doing a normality test to ensure that the underlying PDF is approximately Gaussian. The price to pay, if one wants to maintain the $1-3\times 10^{-7}$ CL, is a detection threshold up to 10 times higher at $1\lambda/D$ than commonly used at larger separations, and up to 2 times higher at $2\lambda/D$. Of course, this penalty puts heavy burden on contrast requirements for very small inner working angle coronagraphs. The penalty decreases rapidly with separation though, for example by a factor of 5 between 1 and 2 $\lambda/D$, which indicates that $2\lambda/D$ might potentially be a practical limit to small inner working angle coronagraphy. Note that this statistical limitation does not preclude detections at very small angles (see, e.g., Mawet et al.~2011\nocite{Mawet2011a}), but makes it significantly harder from a rigorous signal detection theory standpoint.

Alternatively, the same good sensitivity to $5\sigma$ sources can be maintained at all separations, but one has to accept the unavoidable reduced confidence levels, down to $\sim 1-0.00134993$ (equivalent to $3\sigma$ Gaussian), and corresponding increase in false alarm probability at small angles. Even at relatively larger radii, false alarm probabilities can be significantly degraded, e.g., a factor 10 degradation at $r=10\lambda/D$, and a factor 100 at $r=4\lambda/D$! Finally, and most importantly, we argue here that the completeness contrast is in some contexts (surveys) more relevant than the false alarm contrast, justifying lowering the threshold to, e.g.~$\tau=3\sigma$, which reduces relative small-sample penalty factors significantly, and yields for instance a $95\%$ completeness contrast of $\sim 5\sigma$. To conclude this work, we strongly advise the high contrast imaging community to take a deeper look at contrast definitions and speckle statistics in general, and put them in the broader context of signal detection theory. This is particularly relevant in the wake of the golden age of high contrast imaging, with many new systems and upgrades coming online (e.g.~SPHERE \citep{Kasper2012}, GPI \citep{Macintosh2014}, and recent NACO upgrades \citep{Mawet2013,Absil2013}), as well as a series of ground-based extremely large telescope (ELT) projects (e.g.~PFI \citep{Macintosh2006} and EPICS/PCS, see \citet{Kasper2010}), and space-based missions (e.g.~WFIRST-AFTA, see \citet{Spergel2013}, or EXCEDE, see \cite{Guyon2012}).

\acknowledgments

This work was carried out at the European Southern Observatory (ESO) site of Vitacura (Santiago, Chile). The authors would like to thank the referee Prof.~Dmitry Savransky, for his critical, thorough and very constructive review of the manuscript.

\bibliography{smallsample_rev1d}

\end{document}